\documentclass[epj]{svjour}

\usepackage{graphicx}
\usepackage{fancyhdr}
\def\makeheadbox{{
 }}
\def\mytitle{My title} 
\def\myauthors{My name}  
\def\mytype{My type of session}
\def\mysession{My session}

\def\mytitle{Searches for Physics beyond the Standard Model from HERA}
\def\myauthors{Peter Schleper}    
\def\mytype{Contributed Talk}    
\def\mysession{Alternatives}

\begin{document}
\title{Searches for Physics beyond the Standard Model from HERA}
\author{Peter Schleper\inst{1}
\thanks{\emph{Email:} Peter.Schleper@desy.de \newline Proceedings SUSY-2007}%
}                     
%
%
\institute{Hamburg University}
%
\date{}
\abstract{
Searches for physics beyond the Standard Model are reported from
electron-proton collisions at a center of mass energy of 318
GeV. Results on a completely model independent search for deviations from
Standard Model predictions at large transverse momenta are reported as
well as generic searches for contact interactions, leptoquarks  and excited
fermions from the H1 and ZEUS experiments. 
} 
\maketitle
%

\section{Electron-Proton Data from HERA}
\label{HERA}
The HERA collider at DESY has provided  since 1992 electron-proton data
at centre-of-mass (CMS) energies a factor 15 larger than at previous fixed
target experiments, and provides the highest CMS energy for collisions 
with a lepton in the initial state. The HERA II run, which ended in
June 2007, has delivered both electron and positron beams with left and
right handed polarisation of 40\% on average. 
The total integrated luminosity used per experiment is close to 0.5
fb$^{-1}$, an increase by about a factor of four with respect to the HERA I run
alone. First preliminary analyses of the H1 and ZEUS experiments
making use in most cases of all these final data sets will be presented
in the following.  Typical trigger thresholds of 5 to 10 GeV for
electrons and jets have lead to about 10$^8$ triggered events. 
Systematic errors of  measurements are most often dominated by the
1 - 3 \% uncertainty of calorimeter energy scales as obtained from
kinematic constraints. Luminosity and polarisation are measured with
1.6 - 3 \% and 3 - 5\% uncertainty, respectively.

Recent HERA results on Standard Model (SM) processes include 
Neutral and Charged Current cross sections ranging up to Bjorken-$x$
values of 0.65.
The large CMS energy implies measurements of momentum
transfer $Q^2$ up to 30000 GeV$^2$, i.e. far beyond $M_{W,Z}^2$, where
the helicity structure of weak interactions is probed for all
four combinations of electron charge and polarisation, $e^{\pm}_{L,R}$. 
Using inclusive as well as jet data at large transverse momenta has
provided the worlds most precise determinations of parton density
functions and, in a recent combination of H1 and ZEUS data, a
determination of the strong coupling constant from jets of  
$\alpha_s(M_Z) = 0.1198 \pm 0.0019(exp.) \pm 0.0026(th.)$, where the
theory error is dominated by the scale uncertainty of the NLO
calculation.
\section{Model- independent Search}
\label{modelindependent} 
\begin{figure}
\includegraphics[width=0.45\textwidth,angle=0]{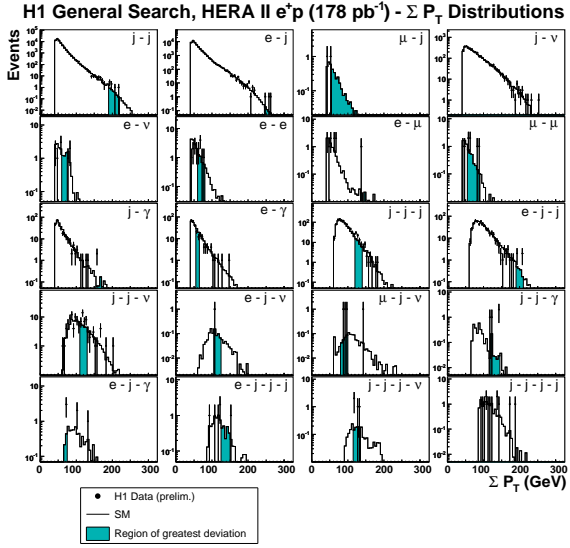}
\caption{Transverse momentum distributions for the event classes of
  the H1 model independent search \cite{H1_inclusive} for $e^+ p$
  data. The shaded   areas denote the regions of maximum deviation
  from SM predictions.} 
\label{fig_inclusive_PT}      
\end{figure}

In comparison to direct searches for specific signatures predicted by
models beyond the SM, generic, model-independent
searches have both advantages and disadvantages. 
In model independent searches 
\begin{itemize}
 \item many final states are
   investigated simultaneously leading to a rather complete view on
   agreement or disagreement with SM predictions;
 \item the phase space for each channel cannot be restricted very much
   in order not to bias the selection to specific assumptions on the
   final state kinematics. Unfortunately this also implies that the
   separation power between a possible signal and SM background is
   significantly reduced; 
 \item even if no deviation from SM predictions is found it is not
   possible to argue that new physics cannot be found in the same
   data, because of the reduced separation power;
 \item any disagreement with respect to SM predictions is difficult to
   quantify, because the likelihood for single, large deviations is
   sizeable once many distributions with many possible regions of
   interest have been looked at. 
\end{itemize}
In a recent analysis  \cite{H1_inclusive} the H1 collaboration
has presented a solution to this last item. Using all HERA II data,
final states with any number of electrons, muons, photons and
hadronic jets with transverse momentum $P_T>20 $ GeV as well as
possible missing transverse momentum $P_T^{miss} > 20$ GeV have been
investigated. The resulting event classes have been analysed with
respect to invariant mass $M$ of the hard process and transverse scalar
momentum $P_T =\sum \left| P_{T,i} \right|$, as motivated by possible
new resonances or Jacobian peaks. For each channel all possible
kinematic ranges of $M$ and $P_T$ have been considered to find the one
with the most significant deviation from SM predictions.
Examples for the $P_T$ distributions of all event classes with data
are shown in Figure \ref{fig_inclusive_PT}.
For the processes with the largest cross sections, i.e. $ej,\, \nu j,
\, jj$, where $j$ denotes jet and  $\nu$ denotes
$P_T^{miss}$, transverse momenta up to 250 GeV are reached. For large
multiplicity classes the number of events is very small, as expected.
The agreement with SM expectations is generally convincing. To
quantify this, all possible combinations of lower and upper cuts on
the quantity shown are applied in order to select the region with
largest statistical difference with respect to expectation for each event
class (shaded areas). 
\begin{figure} 
\includegraphics[width=0.45\textwidth,angle=0]{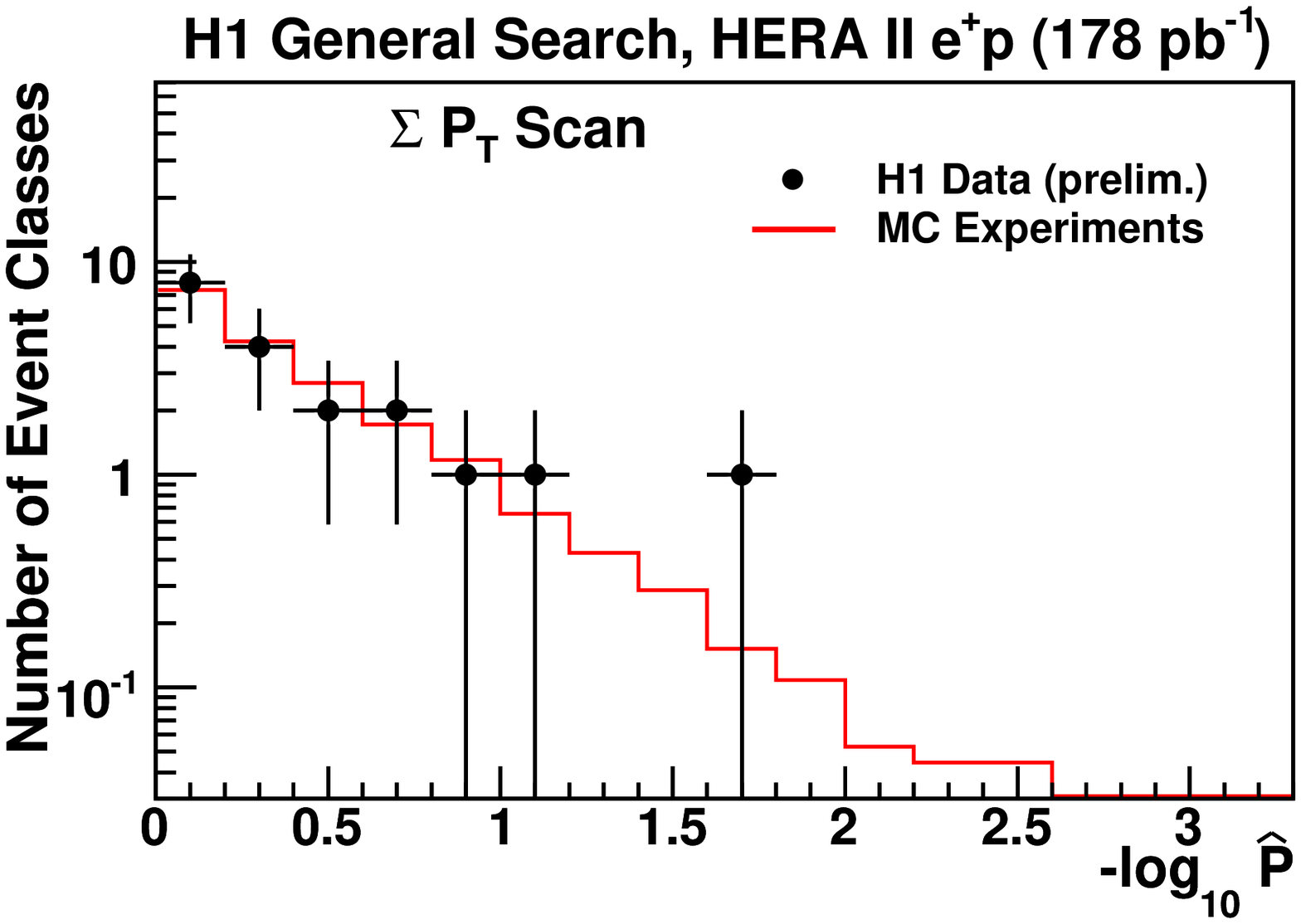} 
\includegraphics[width=0.45\textwidth,angle=0]{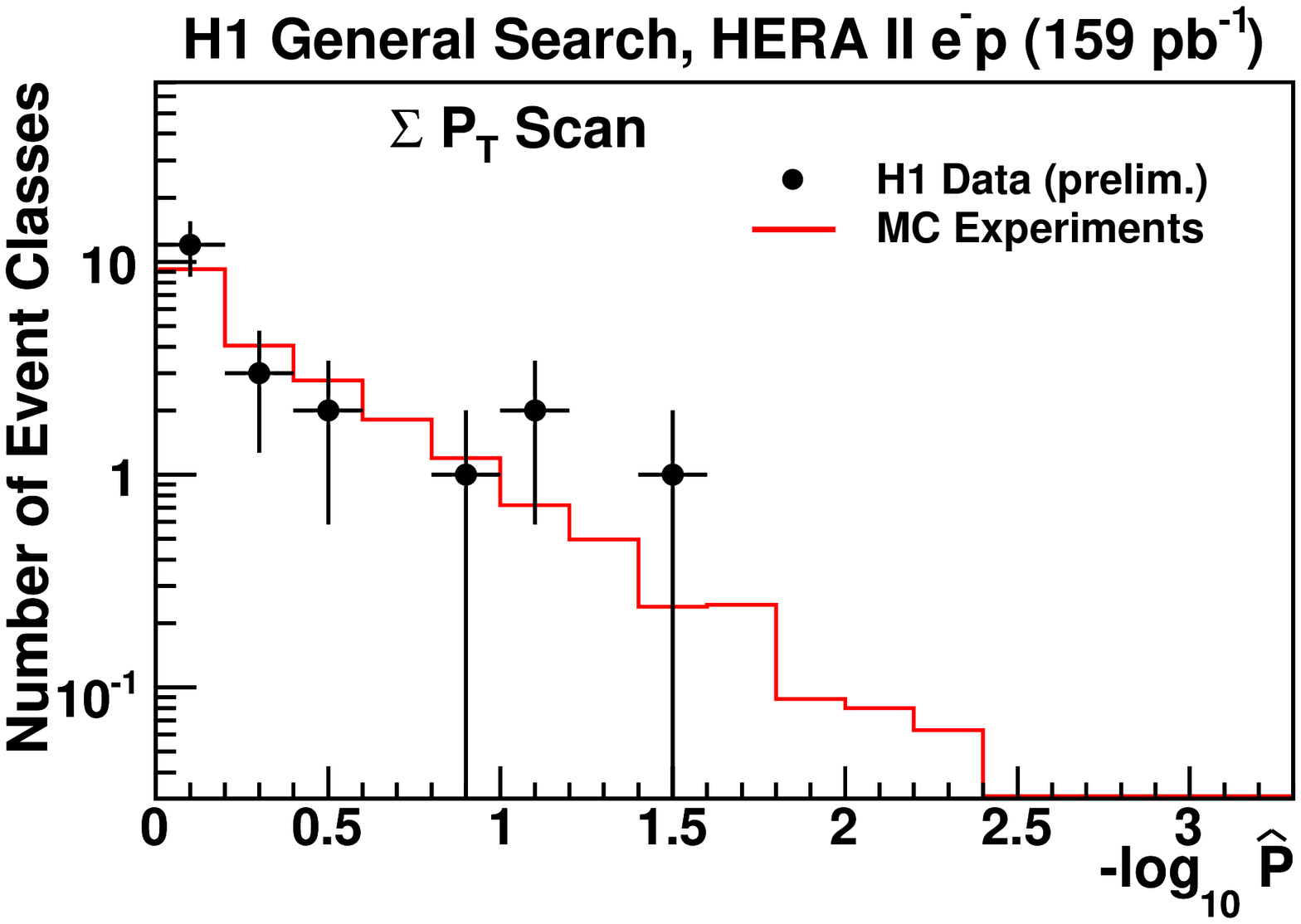} 
\caption{Probability of the maximum deviation within the $P_T$ 
   distribution of all event classes, for $e^+p$
  (top) and $e^-p$ (bottom) data from H1 \cite{H1_inclusive}.}  
\label{fig_prob}
\end{figure}
Taking into account experimental and theoretical errors for both data
and prediction for each event class, the probability $\hat{P}$ for observing a
maximum deviation larger than the one in the data has been determined
from a large number of Monte Carlo experiments, which are based on the SM
prediction. Figure \ref{fig_prob} shows the distribution of these
probabilities for all event classes. The distribution of probabilities
$\hat{P}$ 
for large fluctuations in the data closely follows the expectation of
the Monte Carlo experiments, indicating that overall the data set as
well as the statistical procedures are well understood. For the 88
distributions looked at, the largest
deviation in the data has a probability of about 2\% ($-\log_{10}\hat{P} =
1.7$). It corresponds to final states with a muon, a jet and missing
transverse momentum. A more detailed study of this event class is
presented in \cite{H1_highptlepton}. Note that this deviation is not
observed by the ZEUS experiment in a comparable analysis
\cite{ZEUS_highptlepton}. 
\section{Contact Interactions} 
Extensions of the SM often involve new particles with masses beyond
the kinematic reach of current colliders, which nevertheless imply new
interactions between the known fermions. In this case the 
4-momentum transfer $Q^2$ is
smaller than the mass of the new particle, so that the propagator 
can be approximated by a constant. In this 
``contact interaction'' limit, many types of new interactions can thus
be parameterised by a pure 4-fermion interaction of the form (for $ep$
scattering at HERA) 
$$L_{CI} = \sum_{i,j=L,R;q=u,..b} \eta_{ij}^q \, J^{\mu}_{e,i} \,
J_{\mu\,q,j} $$
where e.g. $J^{\mu}_{e,L} = \bar{e}_L\gamma^{\mu}e_L$ denotes the
common left-handed electron current. While historically contact
interactions have been discussed mostly in the context of new gauge
bosons, compositeness models and fermion radius, it was recently
also used to constrain large extra dimensions and even unparticle
physics \cite{unparticle}. In any of these cases the constants
$\eta_{ij}^q$ have to be 
chosen appropriately to reflect the scale  and the helicity
structure of the new interaction.

At HERA, the high precision data on inclusive
deep inelastic scattering, $ep\rightarrow eX$, as a function of $Q^2$
have been used to 
constrain contact interactions. Here an analysis \cite{ZEUS_CI} by the
ZEUS collaboration is presented which is based on an integrated
luminosity of 274 pb$^{-1}$. The data exhibits the sharp decrease as a 
function of $Q^2$ expected due to the approximate $1/Q^4$ dependence of the
propagator for $\gamma/Z$ interference folded with the steeply falling
quark distributions at large $x$.  
\begin{figure} 
\includegraphics[width=0.45\textwidth,angle=0]{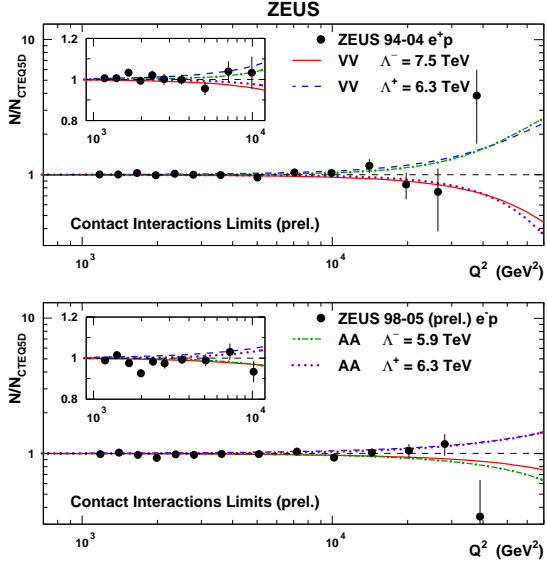} 
\caption{Ratio of inclusive neutral current deep inelastic scattering
   data and SM expectation as a function of $Q^2$, 
   for $e^+p$   (top) and $e^-p$ (bottom) data from ZEUS \cite{ZEUS_CI}.
   Also shown are predictions based on contact interactions with pure
   vector (VV) and pure axial-vector (AA) currents. The scales
   $\Lambda^{\pm}$ chosen here for positive ($+$) or negative
  ($-$) interference correspond to the limits (95\% C.L.) obtained
   by comparing these predictions with the data shown.}
\label{ZEUS_VVAA}
\end{figure}
Shown in Figure \ref{ZEUS_VVAA} is the ratio of
the data and the SM prediction which is based on a NLO QCD fit to
inclusive DIS data from ZEUS and from fixed target experiments. Note
that the prediction used here, in the region of interest at high
$Q^2$, is dominated by the precision of fixed target experiments,
which have measured at large $x$ but much smaller $Q^2$, and of the
$\alpha_s$ dependent QCD evolution of parton densities towards large
$Q^2$. Therefore the bias of using ZEUS for both data and SM prediction
is small.

Within errors, the data show no significant deviation with respect to the
SM expectation.  
Taking into account experimental statistical and systematic
uncertainties as well as the error on $\alpha_s$ and on the parton
densities, fits to the strength of various models leading to contact
interactions have been performed. 
\begin{figure} 
\includegraphics[width=0.45\textwidth,angle=0]{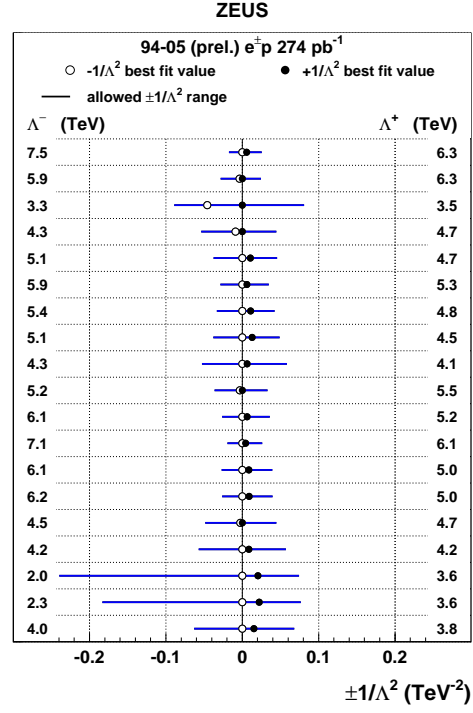} 
\caption{Confidence intervals of $\pm 1/\Lambda^2$ for general contact
    interaction scenarios.
    The numbers at the right (left) margin are the corresponding lower
    limits on the mass scale $\Lambda^+$ 
    ($\Lambda^-$). The filled (open) circles indicate the positions
    corresponding to the best-fit coupling values, for positive
    (negative) couplings \cite{ZEUS_CI}.} 
\label{CI_table}
\end{figure}
\begin{itemize}
  \item Model independent results on contact interactions, depending
    only on the helicity structure, have been determined. In the strong
    coupling limit $\eta = 4\pi / \Lambda^2$, scales $\Lambda$ in the
    range of 2 TeV to 7.5 TeV can be excluded (Figure \ref{CI_table}).  
  \item Leptoquarks (LQ) with masses $M_{LQ}$ above the CMS energy of
    HERA (318 GeV)  
    might contribute to the inclusive $ep$ cross section
    via virtual exchange. At high $M_{LQ}$, the propagator for LQ
    exchange will contract to a form 
    $\eta \sim \lambda^2 / M_{LQ}^2$,
    where the coupling 
    $\lambda$ refers to the $e-q-LQ$ vertex. Considering all possible
    quantum numbers for leptoquarks
    compatible with the gauge symmetries of the SM as well as lepton
    and baryon number conservation, limits on $M_{LQ} / \lambda$ in the range
    0.3 TeV to 1.9 TeV are derived. 
    For masses below 318 GeV these bounds are complemented 
    by direct searches for LQ decays, which lead to
    bounds on $\lambda$ much smaller than $10^{-2}$.
  \item Squarks in R-parity violating Supersymmetry with a
    $\lambda'_{ijk} L_i Q_j D_k$ type interaction are constrained
    (simular to leptoquarks) to $M_{LQ} / \lambda'>0.44$
    TeV.
  \item Large extra dimensions also lead to deviations from SM predictions
    and can thus be constrained to scales $M_S>0.88 $ TeV.
  \item A finite radius $R_q$ of quarks leads to a form factor
    which changes the cross section by a factor $(1 - R_q^2 Q^2 /
    6)^2$. The quark radius can thus be constrained to $R_q < 6.7
    \times 10^{-19}$m. 
\end{itemize}
\section{Search for excited leptons}
\label{excited} 
\begin{figure} 
\includegraphics[width=0.4\textwidth,angle=0]{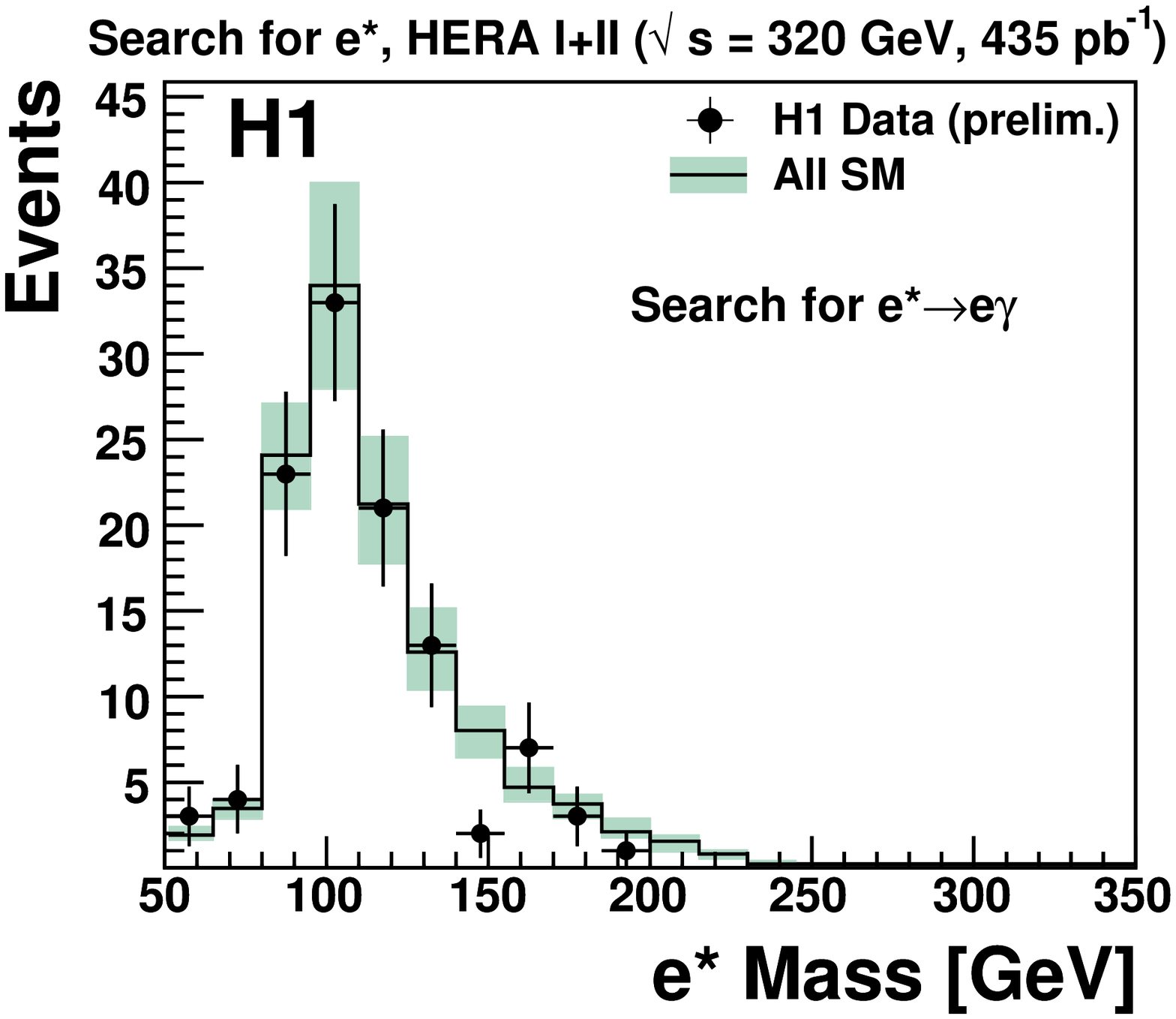} 
\includegraphics[width=0.4\textwidth,angle=0]{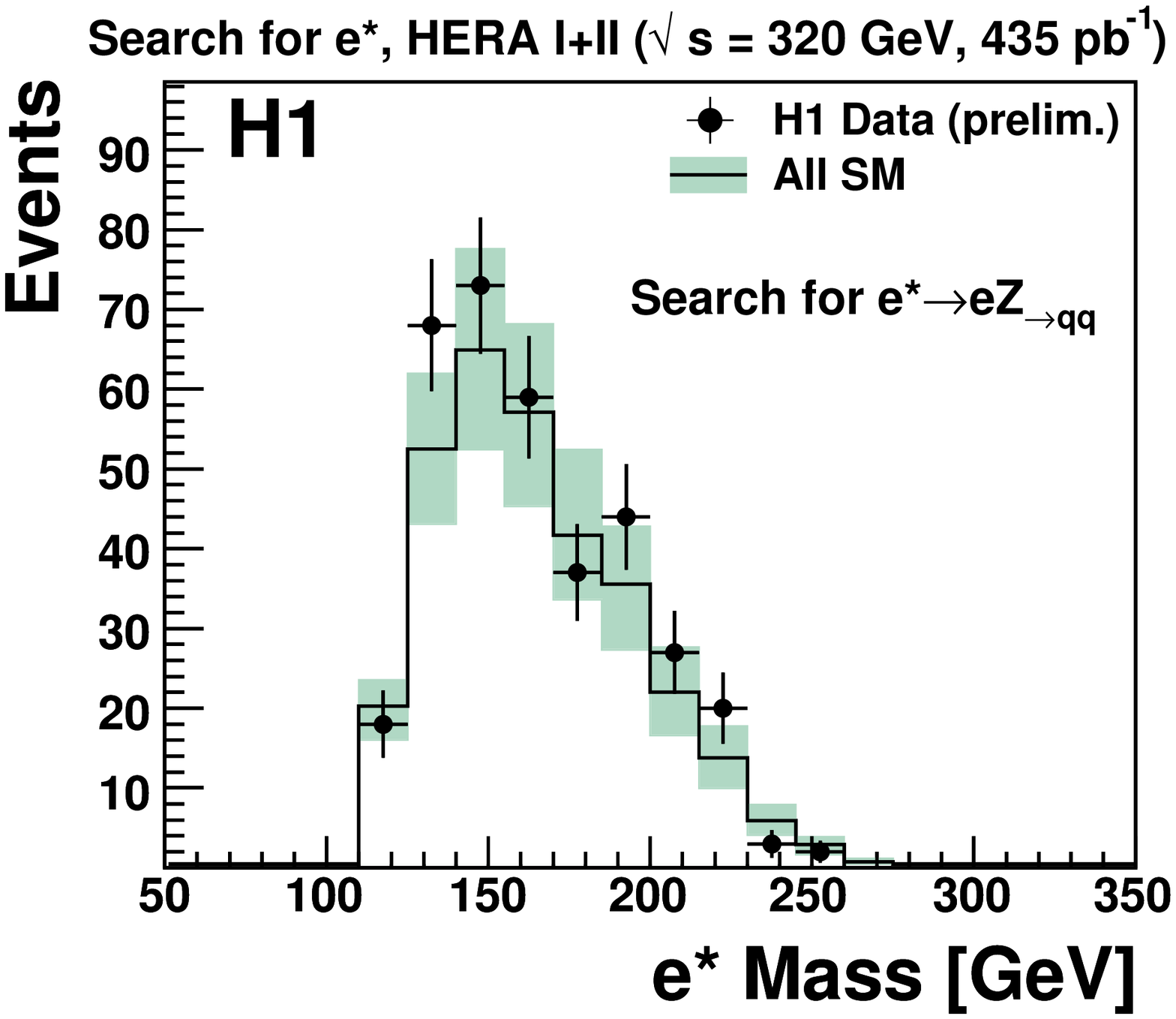} 
\includegraphics[width=0.4\textwidth,angle=0]{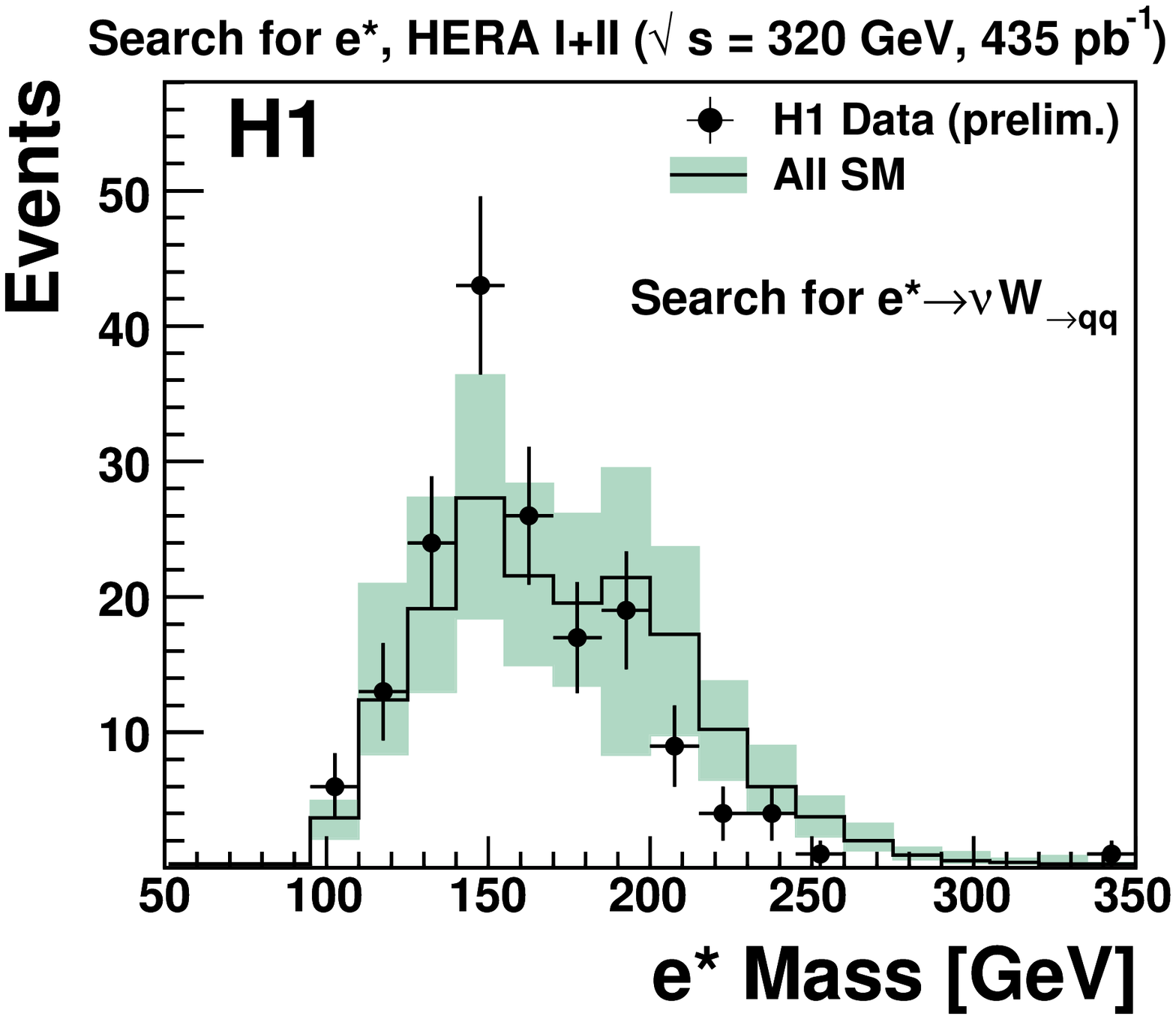} 
\caption{Mass distributions of $e^*$ candidates in final states with
  $\gamma, Z\rightarrow q\bar{q}$ or $W\rightarrow q\bar{q}'$ from H1
   \cite{H1_estar}.} 
\label{H1_estar_masses}
\end{figure}
In models of composite fermions, excited states are to be expected
which can be produced in processes with large momentum transfer. 
For experimental investigations the LEP and HERA experiments have
employed the most general model with spin 1/2, isospin 1/2, left and
right handed symmetric doublets of excited fermions. 
The corresponding term in the lagrangian for the tensor coupling
between a SM lepton doublet, an 
excited lepton and the  SM gauge bosons is
$$ L_{L^*L} = \frac{1}{2\Lambda} \bar{L}^*_R \, \sigma^{\mu\nu}
          \left[ g f \frac{\vec{\tau}}{2} \partial_{\mu} \vec{W}_{\nu}
                +g'f'\frac{Y}{2}          \partial_{\mu}      B_{\nu}
          \right]
           L_L \, +  h.c.
$$
Here $L^*_R$ denotes a righthanded doublet of excited leptons,
$\sigma^{\mu\nu}=i/2\,(\gamma^{\mu}\gamma^{\nu}-\gamma^{\nu}\gamma^{\mu})$, 
$g$ and $g'$ are the SM gauge couplings
and $\Lambda$ denotes the compositeness scale. The factor $f$ thus
describes the strength of the $SU(2)$ vertex
$\nu^* e W^{\pm}$ relative to the
$\nu e W^{\pm}$ vertex of the SM, and similarly $f'$ for $U(1)_Y$ vertices.

Based on all
their data taken at the largest HERA CMS energy of 318 GeV,
the H1 collaboration has presented a preliminary analysis
\cite{H1_estar} for the
decay modes $e^* \rightarrow e\gamma$, $e^* \rightarrow eZ$ and 
$e^*\rightarrow \nu W$ with $W,\,Z$ decaying hadronically.   
\begin{figure} 
\includegraphics[width=0.45\textwidth,angle=0]{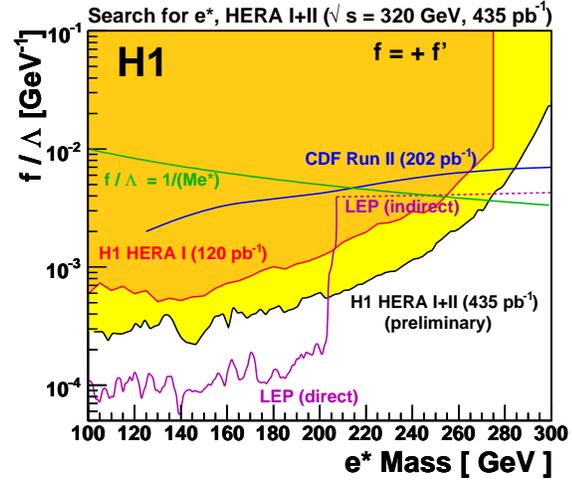} 
\caption{Limits of excited electrons from LEP, Tevatron and
  H1\cite{H1_estar}.}
\label{H1_estar_limits}
\end{figure}
Figure \ref{H1_estar_masses} shows the corresponding reconstructed
masses of $e^*$ candidates in the three decay modes. As no significant
excess above the SM expectation is observed, limits as a function of mass and
coupling $f/\Lambda$ (for $f=f^*$) are derived (Figure
\ref{H1_estar_limits}). To give an example, $e^*$ masses below 260 GeV,
i.e. far beyond the kinematic reach of LEP, are excluded for
$\Lambda$=1 TeV and $f$=2.
\section{Conclusion}
The H1 and ZEUS experiments have searched for deviations from Standard
Model predictions, both in a completely model-independent way by looking in
basically all final states, and in a partially model independent way
by invoking the most general form of effective lagrangians for contact
interactions, leptoquarks, large extra dimensions  and compositeness
(quark radius and excited leptons). In preliminary analyses using in
most cases all their high energy data until the end of the HERA
running in June 2007, no significant 
deviations have been found and constraints on the corresponding
couplings are derived for masses reaching typically up to 280 GeV. It
is a remarkable success that, at the large scales of interest here, 
all the many final state measurements at HERA are correctly described
by Standard Model predictions.
%


\begin{thebibliography}{9}
\bibitem{H1_inclusive}[H1 collaboration]   H1prelim-07-061,
  contributed paper to EPS2007, Abstract 199. 
\bibitem{H1_highptlepton}[H1 collaboration] H1prelim-07-063,
  contributed paper to EPS2007, Abstract 228.
\bibitem{ZEUS_highptlepton}
  [ZEUS collaboration]         ZEUS-prel-07-021, contributed paper to
  EPS2007, Abstract 79; 
  [H1 and ZEUS collaborations] ZEUS-prel-07-029 and H1prelim-07-162, ibid. 
\bibitem{unparticle} G-J. Ding, M-L. Yan, Phys.Rev.D76:075005,2007.
\bibitem{ZEUS_CI}[ZEUS collaboration] 
  ZEUS-prel-06-018, contributed paper to ICHEP06, Abstract 89.
\bibitem{H1_estar}[H1 collaboration] H1prelim-07-065, contributed
  paper to EPS2007, Abstract 201.
\end{thebibliography}
\end{document}